\documentstyle[12pt]{article}
\begin{document}
\def\y{\rule{0.5pt}{7pt}}
\def\Y{\rule[-2.2mm]{0.7pt}{20pt}}
\def\yy{\rule[-1mm]{0.6pt}{12pt}}
\begin{center}
{\large \bf A theorem on topologically massive gravity}\\[15mm]
{\large A. N. Aliev  and  Y. Nutku}\\[2mm]
 T\"{U}B\.{I}TAK - Marmara Research Center \\
 Research Institute for Basic Sciences \\
 Department of Physics \\
 41470 Gebze, Turkey \\[10mm]
\end{center}
\noindent

\vspace{1cm}

  We show that for three dimensional space-times admitting a hypersurface
orthogonal Killing vector field Deser, Jackiw and Templeton's vacuum field
equations of topologically massive gravity allow only the trivial flat
space-time solution. Thus spin is necessary to support topological mass.

\vspace{1cm}

  Deser, Jackiw and Templeton's theory of topologically massive gravity
is the dynamical theory of gravitation in three dimensions \cite{djt}.
The field equations of DJT consist of a proportionality
between the Einstein and Cotton tensors
\begin{equation}
G_{ik} + \frac{1}{\mu} \, C_{ik} =0
\label{djteqs}
\end{equation}
where the constant of proportionality  $ \mu $ is the topological mass.
The fact that the Einstein tensor is of second order, whereas the Cotton
tensor is of third order in the derivatives of the metric
is a major stumbling block in
constructing exact solutions of eqs.(\ref{djteqs}). In fact most of the
known solutions of the DJT field equations are homogeneous spaces \cite{slns}.
In this letter we shall show that there is another severe restriction on
solutions of DJT field equations which is expressed by the following

\vspace{5mm}

\noindent
{\bf Theorem:} {\it If a three dimensional (pseudo-)Riemannian manifold
admits a hypersurface orthogonal Killing vector, then there are no
non-trivial solutions to the vacuum field equations of topologically
massive gravity.}
\vspace{5mm}

   The proof of this theorem follows from a simple exploration of the
consequences of the equations
\begin{equation}
\xi_{(i;j)} = 0,
\label{killing}
\end{equation}
\begin{equation}
\xi_{[i}\, \xi_{j;k]} = 0
\label{killingh}
\end{equation}
governing a hypersurface orthogonal Killing vector.
We recall also the definition of the Riemann tensor
\begin{equation}
\xi_{i;[j;k]} = \frac{1}{2}  \xi_{l} \, R^{l}\,_{ijk}
\label{riemann}
\end{equation}
and the Lie derivative of the Ricci tensor along the vector field $\xi$
\begin{equation}
\pounds_{\xi} R_{ik} = R_{i k; m} \,\, \xi^{m}
 + R_{im} \,\xi^{m}_{\;\;\;\;\; ;k}  + R_{mk} \,\xi^{m}_{\;\;\;\;\; ;i} = 0
\label{lie}
\end{equation}
which vanishes when $\xi$ is Killing. Unlike the Einstein tensor
the Cotton tensor is traceless and therefore for vacuum the DJT field
equations (\ref{djteqs}) imply the vanishing of the curvature scalar.
Thus for vacuum solutions of topologically massive gravity the Einstein
tensor reduces to the Ricci tensor and the definition
of the Cotton tensor  simplifies somewhat
\begin{equation}
 C^{i}_{\;j} = \frac{1}{\sqrt{\yy \, g \, \yy}} \,
 \epsilon^{imn} \, R_{jn;m}
\label{cotton}
\end{equation}
where  $ g $ denotes the determinant of the metric and
 $\epsilon^{ijk}$ is the completely skew Levi-Civita tensor density.

    The basic idea of the proof lies in showing that there is a
complete mismatch between the components of the Einstein and Cotton
tensors in directions parallel and orthogonal to a
hypersurface orthogonal Killing vector.

   We shall denote components along and orthogonal to the Killing vector
by $*$ and $\perp$ respectively. Thus for an arbitrary vector field $A$
$$ A_{*} \equiv A_{i} \, \xi^{i}          \hspace{17mm}
A_{\perp} \equiv A_{i} \, h^{i}_{\;k} $$
where $ h^{i}_{\;k} $ is the projection operator
\begin{equation} \begin{array}{l}
h_{\;j}^{i} = \delta_{\;j}^{i}
    - \frac{\textstyle{ \xi^{i} \xi_{j} }}{\textstyle {\xi^2}} \, , \\[3mm]
h_{\;j}^{i} \, h_{\;k}^{j} = h_{\;k}^{i},\hspace{2cm} h_{\;j}^{i} \, \xi^j =0
\end{array}         \label{proj}
\end{equation}
and $ \xi^2 $ is the square of the magnitude of the Killing vector.
We shall show that eqs.(\ref{killing}) and (\ref{killingh}) imply
\begin{equation} \begin{array}{llllll}
R_{**} & \not\equiv & 0,& \hspace{2cm} C_{**}& \equiv & 0 \\
R_{*\perp}& \equiv & 0,& \hspace{2cm} C_{*\perp}& \not\equiv & 0 \\
R_{\perp\perp} & \not\equiv & 0,& \hspace{2cm} C_{\perp\perp}& \equiv & 0
\end{array}             \label{mismatch}
\end{equation}
which requires that the Ricci as well as the Cotton tensors in
eqs.(\ref{djteqs}) must both vanish simultaneously. Therefore it is
impossible to satisfy eqs.(\ref{djteqs}) in a nontrivial way.

   Two important relations which will be used in the proof
are given by the following
\vspace{2mm}

\noindent
{\bf Lemma:} {\it For a hypersurface orthogonal Killing vector}
\begin{equation}
\xi^{m} R_{m \,[i} \xi_{\,j]}  =  0     ,
\label{lemma}
\end{equation}
\begin{equation}
\xi^{m} R_{m \,[i} \xi_{\,j;k]}  =  0   .
\label{lemma2}
\end{equation}
\vspace{2mm}

\noindent
In order to show that eq.(\ref{lemma}) is a consequence of
eqs.(\ref{killing}) and (\ref{killingh}) we take the covariant
derivative of the hyper\-surface orthogonality condition
$$    \xi_{[i;j} \, \xi_{k];l} +  \xi_{[k} \,  \xi_{i;j];l}  = 0 $$
and using the definition of the Riemann tensor (\ref{riemann})
rewrite it in the form
$$  \xi_{m} \, R^{m}\,_{l[ji}\, \xi_{k]} +  \xi_{[i;j}\, \xi_{k];l} =  0 $$
and finally contract with $ g^{lj} $ to arrive at eq.(\ref{lemma}).
Eq.(\ref{lemma}) will be used repeatedly in the proof of the first two
lines of the mismatch in eqs.(\ref{mismatch}), on the other hand in order
to show that eq.(\ref{lemma2}) is satisfied we need knowledge of
$ R_{*\perp} = 0 $ which will come at a later stage.
However, if we were to assume now the validity of a simple result which
is derived in the second line of eqs.(\ref{riccitr1}) in the proof of
$ R_{*\perp} \equiv 0 $, namely
\begin{equation}
R_{ij} \, \xi^{j} = \frac{1}{\xi^2}
R_{jk} \, \xi^{j} \xi^{k} \xi_{i}
\label{onemli}
\end{equation}
we have
$$ \xi^{m} R_{m \,[k} \;\; \xi_{\,i;j]}  =
\frac{1}{\xi^2}  R_{mn} \, \xi^{m} \xi^{n} \; \xi_{[k}  \xi_{i;j]} = 0 $$
and the proof of eq.(\ref{lemma2}) is immediate.

   Now we turn to the proof of the mismatch indicated in eqs.(\ref{mismatch}).
We shall first show that $  C_{**} \equiv 0 $.
From the definition of the $**$ components and eq.(\ref{cotton}) we have
$$ \begin{array}{lll}         C_{**} & = &
 C_{ik} \, \xi^{i} \xi^{k} = \frac{\textstyle{1}}
 {\textstyle{\sqrt{\yy \, g \, \yy}}}\,
 \epsilon^{ijl} \,R_{kl;j} \,\xi_{i} \, \xi^{k}\\[2mm]
& = & \frac{\textstyle{1}}{\textstyle{\sqrt{\yy \,g \,\yy}}}\left[\epsilon^{ijl}
\left( R_{kl}\, \xi_{i}\, \xi^{k} \right)_{;j}
-\epsilon^{ijl} \,  R_{kl} \left( \xi^{k}_{\,\,;j} \, \xi_{i} +
\xi^{k} \, \xi_{i;j} \right)\right]
\end{array}                  $$
where the first term vanishes by eq.(\ref{lemma}) of the
lemma and the second term can be transformed through the use of the relation
\begin{equation}
\xi_{i;j} = \frac{1}{ \xi^2}  \xi_{[i}\, (\xi^2)_{,j]}
\label{kilasym}
\end{equation}
which follows from the contraction of the hyper-surface orthogonality
condition (\ref{killingh}) with the Killing vector.
As a consequence we have
\begin{equation}
 C_{**}  =
-  \frac{\textstyle{1}}{\textstyle{2\,\sqrt{\yy \,g \, \yy} \,\,\xi^2}} \,
\epsilon^{ijl}
\left[ \right.  \,R^{k}_{l}\, (\xi^2)_{,k} \,\xi_{i}\, \xi_{j}
 +  (\xi^2)_{,i}\, R_{kl} \, \xi^{k}\, \xi_{j}
    - 2 R_{kl}\, \xi^{k} \, \xi_{i}\, (\xi^2)_{,j}\left. \right] = 0
\label{cot**}
\end{equation}
which is readily seen to vanish identically because the first term is
symmetric in $i,j$ and eq.(\ref{lemma}) of the lemma takes care of the last
two terms due to anti-symmetry in $l,j$ and $l,i$ supplied by the
Levi-Civita tensor density in the second and third terms respectively.

In order to get a definite expression for $R_{**}$
we may contract eq.(\ref{riemann}) and use the fact that $\xi$
is a Killing vector to obtain
$$ \xi_{i;k}^{\;\;\;\;\;;k} + R_{ik} \, \xi^{k} = 0 $$
which enables us to write
$$ R_{**}  = R_{ik} \, \xi^{i} \xi^{k}   =
  \xi^{i;k} \, \xi_{i;k} - \frac{1}{2}\,(\xi^2)^{,k}_{\;\;\;;k} $$
and using eq.(\ref{kilasym}) one can transform this expression to the form
$$ R_{**} =  \frac{1}{2 \,\xi^4} \left[ \xi^2 \,
(\xi^2)^{,i}\, (\xi^2)_{,i} - (\xi^{i} \, (\xi^2)_{,i})^2 - \xi^{4}\,
(\xi^2)^{,i}_{\,\,\,;i}\right] $$
where the second term vanishes and we are left with the expression
\begin{equation}
R_{**} =  - \frac{1}{2} \; \xi^2 \left[ \xi^{-2} \, (\xi^2)^{;i} \right]_{;i}
\label{rxx}
\end{equation}
which obviously does not vanish identically.

     For the $ * \perp $ components the argument is very similar.
Starting from the definition of these components it follows that
\begin{equation} \begin{array}{lll}
R_{*\perp} & = & R_{ik} \, \xi^{i} h^{k}_{l}  \\
   & = & R_{il} \, \xi^{i} -
 \frac{\textstyle{1}}{\textstyle{\xi^2}} R_{ik} \, \xi^{i} \xi^{k} \xi_{l}
 \\[3mm]
 & = &  R_{il} \, \xi^{i} -
 \frac{\textstyle{1}}{\textstyle{\xi^2}} R_{il} \, \xi^{i} \xi^{k} \xi_{k}\\
 & \equiv & 0
\end{array}
\label{riccitr1}
\end{equation}
where we have used eq.(\ref{lemma}) of the lemma in next to the last step.
In order to show that $  C_{*\perp}  $ we start from its
definition
$$ \begin{array}{lll}
 C_{*\perp} & = &  C_{ik}\, h^{i}_{j}\, \xi^{k} =
 C_{jk} \, \xi^{k} -   \frac{\textstyle{1}}{\textstyle{\xi^2}} \,
 C_{ik} \, \xi^{i} \, \xi^{k} \, \xi_{j}      \\[4mm]
& = & \frac{\textstyle{1}}{\textstyle{\sqrt{\yy \,g \, \yy}}}
\epsilon^{ijl}  R_{kl;j} \, \xi^{k} =
 \frac{\textstyle{1}}{\textstyle{\sqrt{\yy \,g \, \yy}}}\left[ \epsilon^{ijl}
 \left( R_{kl} \, \xi^{k} \right)_{;j} -
\epsilon^{ijl} \, R_{kl} \,\xi^{k}_{\,\,\,;j} \right]
\end{array} $$
in view of eq.(\ref{cot**}). By means of the relations (\ref{onemli}) and
(\ref{kilasym}) we have
\begin{equation}
 C^{i}_{\,k}\, \xi^{k} = \frac{1}{2 \sqrt{\yy \,g\, \yy}\,\,\xi^4}
 \epsilon^{ijl}  \left[
\,  \left( \xi^2 \right)_{,[l}  \xi_{\,j]} \; R_{**}
\, +  \, \xi_{\,[l} \left( R_{**}\right)_{,j]}  \xi^2
+ \frac{1}{2} \left( \xi^2 \right)^{,k} R_{k[l} \; \xi_{j]} \; \xi^2  \right]
\label{cottrans2}
\end{equation}
which is not identically zero by eq.(\ref{rxx}).

  We shall now turn to the $\perp  \perp $ components of the Cotton and Ricci
tensors  and first show that $ C_{\perp \perp} \equiv 0  $.   That is,
$$  C_{\perp \perp} =  C_{mn}\, h_{i}^{m}\, h_{k}^{n} =
 C_{ik}\,\,-\,\, \frac{C_{mi}\, \xi^{m} \,
\xi_{k}}{\xi^2} \,\,-\,\, \frac{C_{mk}\, \xi^{m}\, \xi_{i}}{\xi^2} $$
which will vanish identically provided that
\begin{equation}
R_{\,k \, [m\,; n \,]} \, \xi^{k} \,\xi_{i} \,\,+  \,\,
R_{\,k\,i \, ; [\,m} \, \xi_{\,n]}\, \xi^{k}  \,\, +     \,\,
 \xi_{\,[\, m} R_{\,n\,] \,i {\,;k}}  \, \xi^{k} = 0
\label{cotperp1}
\end{equation}
where we have used the definition of the Cotton tensor.
In order to show this, first  we take the covariant derivative of
eq.(\ref{lemma}) in the lemma and use the hypersurface orthogonality
condition (\ref{killingh})
$$ \begin{array}{lll}
R_{k m{\,; n}} \, \xi^{k}\, \xi_{i} & = &
R_{ik {\,;n}} \, \,\xi^{k}\, \xi_{m} \,\,+ \,\,
R_{ik}\, \left(\xi_{m}^{\,\,\,;k} \, \xi_{n} \,\, + \,\,2\,\xi^{k}_{\,\,\,;n}
\, \xi_{m}\right)  \\[3mm]
& & - R_{k m} \left( \xi^{k}_{\,\,\,;i} \, \xi_{n}\,+ \,2 \,\xi_{i;\,n} \,
\xi^{k} \right)
\end{array} $$
which reduces to
$$ \begin{array}{lll}
R_{k m{\,; n}} \, \xi^{k}\, \xi_{i} &  = &
R_{ik {\,; n}} \, \xi^{k}\, \xi_{m} \,\,+ \,\,R_{im;k} \xi^k \xi_n \\[3mm]
& & + 2\,\left( R_{ik} \xi^{k}_{\;\;\;;n} \, \xi_{m} \,\,- \,\,
R_{mk} \,\xi^{k} \,\xi_{i\,;\,n}   \right)
\end{array} $$
when we use eq.(\ref{lie}), that the Lie derivative of the Ricci
tensor with respect to the Killing vector vanishes.
Finally, antisymmetrizing this expression in
$ \,\, m \,\, $ and $ \,\,n \,\,$ we obtain
\begin{equation}
R_{k\, [\, m;\, n \,]} \, \xi^{k}\, \xi_{i} \,\,+ \,\,
R_{ki\, ; [m} \, \xi_{n \,]}\, \xi^{k}  \,\, +      \,\,
 \xi_{\,[ \, m}\, R_{n\,]\, i {\,;k}}  \, \xi^{k} \,\,=  \,\,
 3\,  \xi^{k}\, R_{\,k\,[\,i}\, \xi_{m \,{\, ;n}\,]}  = 0
\label{cotperp2}
\end{equation}
where the left-hand side is preciely the same
as that of eq. (\ref{cotperp1}), while the right-hand side
vanishes by eq.(\ref{lemma2}) of the lemma.
The only remaining step is to show that the $ \,\, \perp \perp \,\, $
components of the Ricci tensor are not identically zero. Now
$$ R_{\perp \perp}= R_{mn}\, h_{i}^{m}\, h_{k}^{n} =
\, R_{ik}\,\, - \,\,\frac{\textstyle{1}}{\textstyle{\xi^2}}
\,\, R_{mi} \,\xi^{m} \,\xi_{k} $$
where we have again used the above lemma. Taking into account eqs.(\ref{rxx})
and (\ref{riccitr1})  we can re-express this relation as
\begin{equation}
R_{\perp \perp}= R_{ik}\,\, +\,\, \frac{\textstyle{1}}{\textstyle{2\xi^2}}
\,\,\xi_{i} \, \xi_{k}\, \left(\frac{(\xi^2)_{\,,j}}{\xi^2}\right)^{\,;j}
\end{equation}
which does not vanish identically.

    We have shown that for a hypersurface orthogonal Killing vector
there is a complete mismatch between the components of the Ricci and
Cotton tensors in directions parallel and orthogonal to the Killing
vector field. This mismatch (\ref{mismatch}) requires that
both the Ricci and Cotton tensors must vanish seperately
and results in flat spacetime as the unique solution of vacuum
field equations of topologically massive gravity with a hypersurface
orthogonal Killing vector.

\end{document}